\documentclass[twocolumn]{aastex631}

\shorttitle{The Venus Benchmark for Exoplanet Atmospheres}
\shortauthors{Stephen R. Kane et al.}

\begin{document}

\title{The Venus Benchmark: Resolving Degeneracies in Terrestrial
  Exoplanet Spectra with DAVINCI}

\author[0000-0002-7084-0529]{Stephen R. Kane}
\affiliation{Department of Earth and Planetary Sciences, University of
  California, Riverside, CA 92521, USA}
\email{skane@ucr.edu}

\author[0009-0008-1437-2485]{David S. Alexander}
\affiliation{NASA Goddard Space Flight Center, 8800 Greenbelt Road, Greenbelt, MD 20771, USA}

\author[0000-0002-9581-647X]{Shahid Aslam}
\affiliation{NASA Goddard Space Flight Center, 8800 Greenbelt Road, Greenbelt, MD 20771, USA}

\author[0000-0001-6285-267X]{Giada N. Arney}
\affiliation{NASA Goddard Space Flight Center, 8800 Greenbelt Road, Greenbelt, MD 20771, USA}

\author[0000-0003-1606-5645]{James B. Garvin}
\affiliation{NASA Goddard Space Flight Center, 8800 Greenbelt Road, Greenbelt, MD 20771, USA}

\author[0000-0003-4204-9534]{Stephanie A. Getty}
\affiliation{NASA Goddard Space Flight Center, 8800 Greenbelt Road, Greenbelt, MD 20771, USA}

\author[0000-0001-6869-5118]{Amy E. Hofmann}
\affiliation{Jet Propulsion Laboratory, California Institute of Technology, Pasadena, CA 91109, USA}

\author[0000-0003-1629-6487]{Noam R. Izenberg}
\affiliation{Johns Hopkins University Applied Physics Laboratory, Laurel, MD 20723, USA}

\author[0000-0001-6357-1255]{Natasha M. Johnson}
\affiliation{NASA Goddard Space Flight Center, 8800 Greenbelt Road, Greenbelt, MD 20771, USA}

\author[0000-0002-0000-199X]{Erika Kohler}
\affiliation{NASA Goddard Space Flight Center, 8800 Greenbelt Road, Greenbelt, MD 20771, USA}

\author[0009-0006-9233-1481]{Emma L. Miles}
\affiliation{Department of Earth and Planetary Sciences, University of California, Riverside, CA 92521, USA}


\begin{abstract}

The characterization of terrestrial exoplanets with thick,
CO$_2$-dominated atmospheres via transmission and emission
spectroscopy is limited by degeneracies between atmospheric
composition, cloud structure, and surface conditions. These
degeneracies are particularly acute for planets in or near the Venus
Zone, where sulfuric acid aerosol layers and deep-atmosphere opacity
render the lower atmosphere and surface spectroscopically inaccessible
via many optical/IR wavelengths. Venus provides the only accessible
analog to such worlds, yet current knowledge of its full atmospheric
profile relies on decades-old in situ data with known
limitations. Here we demonstrate that the forthcoming Deep Atmosphere
Venus Investigation of Noble gases, Chemistry, and Imaging (DAVINCI)
mission will provide the comprehensive, high-fidelity atmospheric
profile data necessary to resolve many of these degeneracies. We
quantify how atmospheric profile uncertainties translate into
uncertainties in modeled transmission and thermal emission spectra of
CO$_2$-dominated terrestrial worlds, and show that the spread in the
modeled Venus benchmark spectrum arising from current atmospheric
profile uncertainties will decrease by factors of $\sim$4--5 for
transmission and $\sim$5--15$\times$ for thermal emission after
DAVINCI, providing a correspondingly improved benchmark and prior for
the modeling of exoplanets with Venus-like atmospheres. We further
demonstrate that these improvement factors are larger for the
cloud-free atmospheric case, where the opacity floor is set by
gas-phase processes rather than aerosols, because the pre-DAVINCI data
are most discrepant in the sub-cloud region. Our results highlight the
importance of Venus for the atmospheric characterization of rocky
exoplanets in the JWST and HWO era, and demonstrate how DAVINCI
measurements will directly improve exoplanet atmospheric modeling.

\end{abstract}

\keywords{astrobiology --- planetary systems}


\section{Introduction}
\label{sec:intro}

The detection and characterization of terrestrial exoplanet
atmospheres represents one of the most compelling scientific endeavors
within the exoplanet community
\citep{seager2010,madhusudhan2019,kane2021e}. The James Webb Space
Telescope (JWST) has demonstrated the sensitivity required to detect
molecular features at the level of tens of parts per million in the
transmission spectra of rocky worlds
\citep{lustigyaeger2023b,moran2023,rigby2023a}. Looking ahead, the
Habitable Worlds Observatory (HWO), recommended by the Astro2020
Decadal Survey, will extend these capabilities to the direct imaging
and spectroscopy of Earth-sized planets orbiting Sun-like stars
\citep{harada2024b,kane2024d,stark2024b,harada2025,tuchow2025a}. A
prerequisite for these endeavors is the ability to reliably interpret
the spectra of thick, CO$_2$-dominated atmospheres, a class that may
be among the most common outcomes of terrestrial planet evolution
\citep{wordsworth2013a,kane2014e,ostberg2023a}.

The challenge of interpreting such spectra is fundamentally one of
degeneracy. Transmission spectroscopy probes only a narrow annulus of
the atmosphere at the planetary limb, sampling approximately four
scale heights at each wavelength
\citep{seager2000b,brown2001c,lecavelierdesetangs2008b}. For thick,
cloud-enshrouded atmospheres, aerosol layers truncate the observable
atmospheric column, obscuring molecular features and mimicking the
spectral signatures of atmospheres with substantially different
compositions \citep{kreidberg2016,sing2016,wakeford2019}. In
CO$_2$-dominated atmospheres analogous to that of Venus, the sulfuric
acid (H$_2$SO$_4$) cloud and haze layers extending from roughly 48 to
70~km altitude impose an effective floor on transmission spectroscopy,
rendering the lower $\sim$70\% of the atmospheric column
spectroscopically inaccessible \citep{ehrenreich2012a,barstow2016a}.
The result is a profound degeneracy: spectral features observed above
the cloud deck can be reproduced by a wide range of underlying
atmospheric structures and surface conditions
\citep{benneke2012,griffith2014,lustigyaeger2019b,ostberg2023c}.

Venus is the only terrestrial planet with a thick, CO$_2$-dominated
atmosphere accessible to in situ investigation, and it therefore
serves as an indispensable ground truth for the validation of
exoplanet atmospheric models \citep{kane2019d,kane2024b}. The Venus
Zone (VZ) framework \citep{kane2014e} identifies the orbital region
where a rocky planet may experience a runaway greenhouse, and recent
demographic analyses demonstrate that terrestrial planets in the VZ
are at least as common as those in the habitable zone (HZ)
\citep{kane2014e,kane2016c,ostberg2023a}. The characterization of VZ
worlds is a major science driver for both JWST and HWO, and the
ability to distinguish Venus-like from Earth-like atmospheres is
central to the assessment of planetary habitability
\citep{lustigyaeger2019a,kane2024b,kane2026a}. Yet our knowledge of
the full Venusian atmospheric profile remains remarkably incomplete:
the most detailed in situ measurements of the deep atmosphere were
obtained by the Soviet Venera/Vega probes and the U.S. Pioneer Venus
multiprobe in the 1970s and 1980s \citep{seiff1985,marov2018}, with
limited precision, sparse vertical sampling, and known calibration
issues \citep{marcq2018,taylor2018}.

The Deep Atmosphere Venus Investigation of Noble gases, Chemistry, and
Imaging (DAVINCI) mission, selected by NASA as a Discovery-class
mission in 2021, will address precisely these gaps
\citep{garvin2022}. DAVINCI will deploy a descent sphere through the
full vertical extent of the Venusian atmosphere, from the upper cloud
deck at $\sim$70~km to the near-surface, providing high-precision
measurements of temperature, pressure, composition, and wind structure
at vertical resolutions far exceeding any previous mission. In this
paper, we present a quantitative assessment of how DAVINCI
measurements will resolve the deep-atmosphere degeneracies that
currently limit the interpretation of CO$_2$-dominated terrestrial
exoplanet spectra. In Section~\ref{sec:degen}, we describe the nature
of these degeneracies, Section~\ref{sec:venus} reviews the current
state of Venus atmospheric knowledge, and Section~\ref{sec:davinci}
describes the DAVINCI measurement capabilities relevant to exoplanet
science. Section~\ref{sec:analysis} derives an analytical error
propagation framework quantifying how DAVINCI measurement improvements
translate into improved spectral precision, and explicitly connects
the results to each of the four degeneracies. We discuss implications
for exoplanet modeling, Venus Zone demographics, mission synergies,
and HWO target selection in Section~\ref{sec:disc}, and provide
concluding remarks in Section~\ref{sec:con}.


\section{The Deep-Atmosphere Degeneracy Problem}
\label{sec:degen}


\subsection{Transmission Spectroscopy of Thick Atmospheres}
\label{trans}

Transmission spectroscopy measures the wavelength-dependent transit
depth of a transiting exoplanet, which increases at wavelengths where
atmospheric constituents absorb starlight passing through the
planetary limb \citep{seager2000b,brown2001c}. The amplitude of
spectral features depends on the atmospheric scale height,
\begin{equation}
  H = \frac{k_B T}{\mu m_u g}
  \label{eq:scaleheight}
\end{equation}
where $k_B$ is the Boltzmann constant, $T$ is the atmospheric
temperature, $\mu$ is the mean molecular weight in atomic mass units,
$m_u$ is the atomic mass unit, and $g$ is the surface gravity. For a
CO$_2$-dominated atmosphere ($\mu \approx 44$), the scale height is a
factor of $\sim$19 smaller than for a solar-composition
H$_2$-dominated atmosphere ($\mu \approx 2.3$) at the same
temperature, producing correspondingly weaker spectral features. The
transit depth variation $\Delta \delta$ produced by an absorbing
species then scales as
\begin{equation}
  \Delta \delta \approx \frac{2 R_p N_H H}{R_\star^2}
  \label{eq:transitdepth}
\end{equation}
where $R_p$ is the planetary radius, $R_\star$ is the stellar radius,
and $N_H$ is the number of scale heights over which the species is
optically thick \citep{lecavelierdesetangs2008b}.

\begin{figure*}
  \includegraphics[angle=270,width=\linewidth]{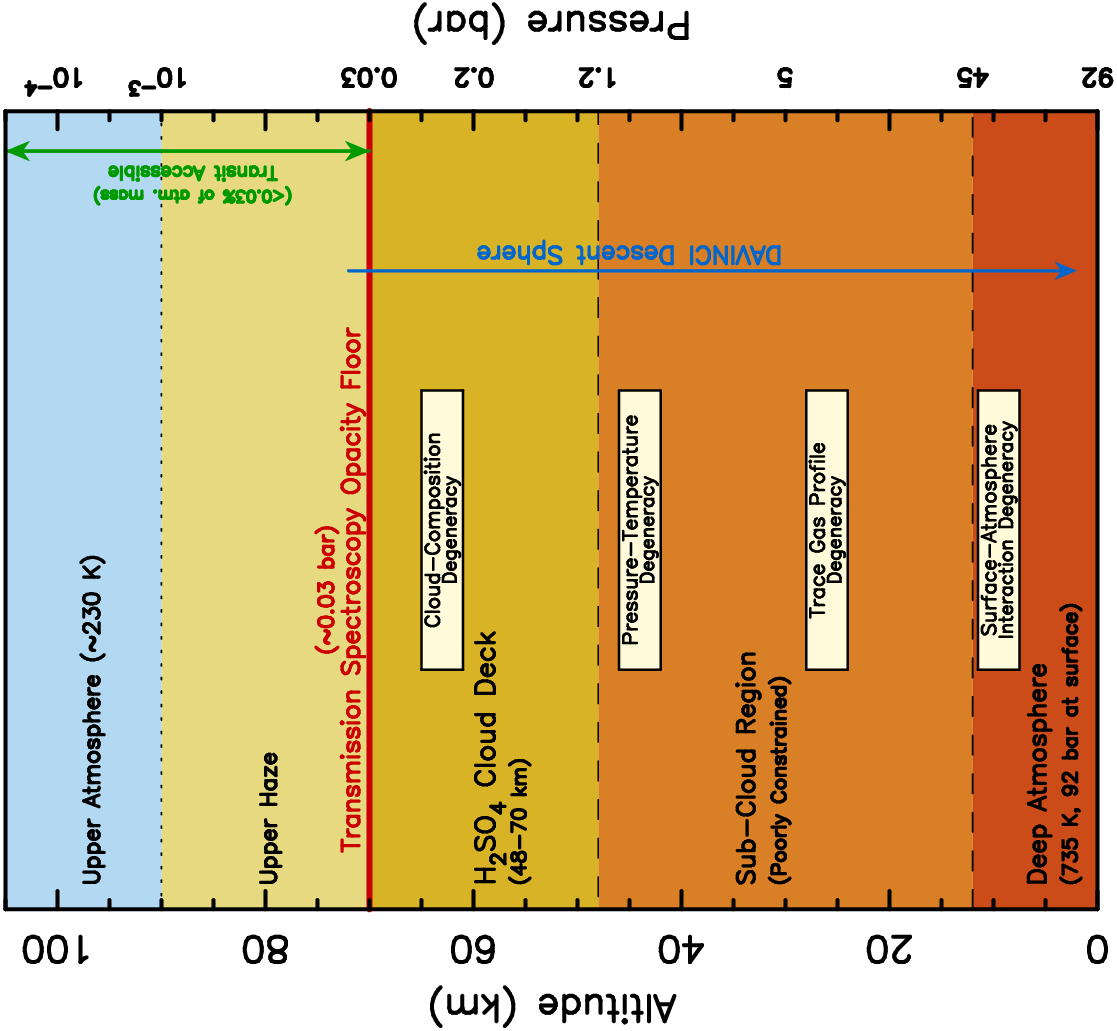}
  \caption{Schematic of the Venus atmospheric structure and its
    implications for exoplanet spectroscopic characterization.  Shown
    altitude-dependent regions are as follows: the deep atmosphere
    (0--12~km, dark orange), the sub-cloud region (12--48~km, medium
    orange), the H$_2$SO$_4$ cloud deck (48--70~km, gold), the upper
    haze (70--90~km, pale yellow), and the upper atmosphere ($>$90~km,
    pale blue).  Atmospheric boundaries are from the Venus
    International Reference Atmosphere \citep{seiff1985} and the cloud
    structure of \citet{knollenberg1980a}. The red line at $\sim$70~km
    marks the transmission spectroscopy opacity floor set by the
    H$_2$SO$_4$ haze layer \citep{ehrenreich2012a,barstow2016a}. This
    corresponds to a pressure of $\sim$0.03~bar, meaning that less
    than 0.03\% of the total atmospheric mass is spectroscopically
    accessible. The blue arrow indicates the DAVINCI descent sphere
    trajectory from the upper cloud deck to the near-surface. The four
    labeled boxes identify the degeneracies described in
    Section~\ref{sources}, located at the altitudes where each
    degeneracy originates. The approximate pressure scale on the right
    axis is derived from the VIRA profile
    \citep{seiff1985,taylor2018}.}
  \label{fig:venatm}
\end{figure*}

The situation is further complicated for Venus-like atmospheres by the
global H$_2$SO$_4$ cloud and haze layer. On Venus, the main cloud deck
extends from $\sim$48 to 70~km, with an upper haze reaching
$\sim$90~km \citep{knollenberg1980a,titov2007,taylor2018}. The cloud
deck is optically thick across a broad wavelength range, setting a
floor below which transmission spectroscopy cannot readily probe. For
a Venus analog observed in transit, the H$_2$SO$_4$ aerosol optical
depth reaches $\tau \sim 1$ at limb geometry near the cloud-top
altitude of $\sim$70~km \citep{ehrenreich2012a,barstow2016a},
corresponding to a pressure of $\sim$0.03~bar in the Venus
International Reference Atmosphere \citep{seiff1985}. Only the
atmosphere above this transmission spectroscopy opacity floor is
spectroscopically accessible, representing less than 0.03\% of the
total atmospheric mass
\citep{ehrenreich2012a}. Figure~\ref{fig:venatm} illustrates this
atmospheric structure, the location of the opacity floor, and the
altitude regions where each of the four degeneracies described below
originates. The figure also shows the DAVINCI descent sphere
trajectory, which will traverse the full atmospheric column from the
upper cloud deck to the near-surface, sampling precisely the regions
that are inaccessible to remote spectroscopic observation.


\subsection{The Cloud-Free Limit}
\label{cloudfree}

The opacity floor at $\sim$70~km shown in Figure~\ref{fig:venatm} is
set by H$_2$SO$_4$ aerosol scattering, but it is important to consider
what would happen in the absence of such a cloud layer. If the
H$_2$SO$_4$ haze were removed entirely, the opacity floor would drop
substantially but would not reach the surface. Two gas-phase effects
impose a lower limit on the altitude accessible to transmission
spectroscopy even in a clear CO$_2$ atmosphere.

First, CO$_2$ Rayleigh scattering and collision-induced absorption
(CIA) become significant at limb geometry due to the enormous path
lengths through a thick atmosphere. Rayleigh scattering cross-sections
scale as $\lambda^{-4}$
\citep{lecavelierdesetangs2008a,lecavelierdesetangs2008b} and thus
dominate at visible and near-ultraviolet wavelengths, while
CO$_2$--CO$_2$ CIA contributes broad continuum opacity across the
infrared \citep{baranov1999,wordsworth2010b,karman2019,tran2024c}.
\citet{wordsworth2010b} showed that CIA becomes a major opacity source
in dense CO$_2$ atmospheres, particularly at pressures exceeding
$\sim$1~bar. Together, these processes would produce a gas-phase
opacity floor at roughly 35--40~km altitude in a 92~bar CO$_2$
atmosphere, depending on wavelength
\citep{ehrenreich2012a,barstow2016a}.

Second, atmospheric refraction limits the depth to which transmission
spectroscopy can probe. In a thick atmosphere, the gradient in
refractive index bends stellar rays passing through the planetary
limb, and below a critical altitude the rays are deflected entirely
out of the observer's beam \citep{misra2014b,betremieux2015}. For a
92~bar CO$_2$ atmosphere, the refraction limit falls at approximately
30--35~km, corresponding to pressures of several
bar. \citet{misra2014c} demonstrated that refraction effects can, in
principle, be used to discriminate between cloudy and clear
atmospheres, but the key point for the present analysis is that even
the optimistic cloud-free case leaves the lowest $\sim$35~km of the
atmosphere and the surface entirely inaccessible to transit-related
spectroscopy.

The distinction between the cloudy and cloud-free opacity floors has
two important implications. First, the deep-atmosphere degeneracies
described in Section~\ref{sources} (particularly the trace gas
profile, pressure-temperature, and surface-atmosphere interaction
degeneracies) persist regardless of aerosol properties. In situ ground
truth from DAVINCI is therefore necessary even for hypothetical
cloud-free CO$_2$-dominated exoplanets. Second, for the realistic case
in which the presence or absence of a Venus-like cloud layer is not
known a priori, the difference between an opacity floor at $\sim$70~km
and one at $\sim$35~km is itself an additional degeneracy, in that a
transit spectrum consistent with a cloudy atmosphere at one
composition may also be consistent with a clear atmosphere at a
different composition. \citet{jordan2021} explored the photochemistry
and haze formation of Venus-like atmospheres around K- and M-dwarf
stars and showed that the aerosol properties, and thus the altitude
of the opacity floor, depend sensitively on spectral type and
ultraviolet irradiation environment. DAVINCI-calibrated models of the
Venus cloud and haze structure will provide the anchor point needed to
disentangle these effects for exoplanets across a range of stellar
environments. We quantify the improvement factors at both opacity
floors in Section~\ref{improvement} and Figure~\ref{fig:venimp}.


\subsection{Sources of Degeneracy}
\label{sources}

The restriction of observations to the upper atmosphere introduces
several distinct but interrelated degeneracies, illustrated
schematically in Figure~\ref{fig:venatm}.

The {\em cloud-composition degeneracy} arises because the spectral
effects of increased aerosol opacity and decreased gas-phase mixing
ratios are partially interchangeable. A muted absorption feature can
result either from a low abundance of the absorbing species or from a
cloud deck that truncates the atmospheric column
\citep{benneke2012,line2013a,sing2016}. For CO$_2$-dominated
atmospheres, this is particularly severe because CO$_2$ is
simultaneously the dominant absorber and the background gas
\citep{lustigyaeger2019b}.

The {\em pressure-temperature degeneracy} stems from the dependence
of both the scale height and molecular absorption cross-sections on
temperature and pressure. Observations confined to the upper
atmosphere cannot independently constrain the deep-atmosphere
temperature lapse rate, the surface temperature, or the surface
pressure \citep{griffith2014,barstow2020a}. On Venus, the temperature
increases from $\sim$230~K at the cloud tops to $\sim$735~K at the
surface, a range that profoundly affects deep-atmosphere chemistry and
radiative transfer but that is invisible to remote spectroscopic
observation.

The {\em trace gas profile degeneracy} concerns the vertical
distributions of chemically active species. On Venus, SO$_2$, OCS,
CO, HCl, and H$_2$O exhibit complex altitude-dependent profiles driven
by photochemistry above the clouds, thermochemical equilibrium in the
deep atmosphere, and heterogeneous reactions within the cloud layer
\citep{krasnopolsky2012a,bierson2020}. The vertical gradients below
the clouds are poorly constrained, yet they are critical inputs to
radiative transfer models that propagate directly into retrieval
results for any CO$_2$-dominated world.

The {\em surface-atmosphere interaction degeneracy} reflects the
influence of surface mineralogy and redox state on atmospheric
composition through weathering, outgassing, and buffering reactions
\citep{fegley1997a,zolotov2018b}. Near-surface abundances of SO$_2$,
OCS, and CO are diagnostic of surface--atmosphere equilibrium, but
remain effectively unknown at the precision required to constrain
surface conditions. For exoplanets, inferred surface conditions are
therefore sensitive to deep-atmosphere chemistry assumptions that
cannot be validated without in situ data from an analog world
\citep{kane2024b}.


\section{Current Venus Atmospheric Knowledge}
\label{sec:venus}


\subsection{Historical In Situ Measurements}
\label{historical}

The in situ exploration of the Venusian atmosphere began with the
Soviet Venera program and the U.S. Pioneer Venus mission. The Venera~4
through 14 landers and the Pioneer Venus multiprobe collectively
provided the foundational measurements that remain the primary
constraints on the deep atmosphere \citep{seiff1985,marov2018}. These
established the basic structure: a predominantly CO$_2$ ($\sim$96.5\%)
and N$_2$ ($\sim$3.5\%) composition, surface conditions of $\sim$735~K
and $\sim$92~bar, a well-defined H$_2$SO$_4$ cloud deck between
$\sim$48 and 70~km, and a suite of photochemically active trace
species \citep{taylor2018}. However, the measurements suffer from
limited mass resolution and dynamic range in the mass spectrometers,
sparse vertical sampling (most probes obtained only a few discrete
composition measurements during descent), and calibration issues in
the extreme lower-atmosphere environment
\citep{hoffman1980d}. Reported values for H$_2$O and SO$_2$ below the
clouds span more than an order of magnitude \citep{marcq2018}.


\subsection{Remote Sensing Constraints}
\label{remote}

ESA's Venus Express mission (2006--2014) substantially improved
understanding of the upper atmosphere and cloud layer
\citep{svedhem2007b,taylor2018}. Near-infrared spectral windows in
night-side emission have been used to probe the lower atmosphere,
constraining CO, H$_2$O, OCS, SO$_2$, and HCl between the surface and
$\sim$40~km \citep{bezard1990b,marcq2018}. However, these windows are
narrow, provide limited vertical resolution, and are sensitive to the
assumed cloud model. Our knowledge of the deep atmosphere therefore
remains dependent on legacy in situ data, supplemented by remote
retrievals subject to the same degeneracies described in
Section~\ref{sources}.


\subsection{Consequences for Exoplanet Models}
\label{consequences}

Contemporary forward models for terrestrial exoplanet spectra require
complete pressure-temperature-composition profiles spanning the full
atmospheric column. Tools such as the Planetary Spectrum Generator
(PSG) \citep{villanueva2018b}, SMART \citep{meadows1996}, and
petitRADTRANS \citep{molliere2019b} assemble Venus-like profiles from
a patchwork of historical data, remote sensing, and photochemical
models, with substantial interpolation in the poorly constrained
0--48~km altitude range \citep{arney2014,lincowski2018}. Uncertainties
propagate through the temperature profile (controlling thermal
emission and scale height), trace gas vertical distributions
(determining spectral features used to distinguish Venus-like from
Earth-like atmospheres; \citealt{lustigyaeger2019b,jordan2021}), and
surface pressure and temperature (essential for climate state
assessment but unconstrained by transit observations alone).


\section{DAVINCI Measurement Capabilities}
\label{sec:davinci}


\subsection{Mission Architecture}
\label{architecture}

DAVINCI comprises a Carrier, Relay, and Imaging Spacecraft (CRIS) and
a Descent Sphere (DS) probe \citep{garvin2022}. The CRIS performs two
Venus flybys for ultraviolet and near-infrared remote sensing before
deploying the DS for atmospheric entry, descent, and near-surface
science over Alpha Regio. The $\sim$1~hour descent provides continuous
measurements from the upper cloud deck at $\sim$70~km to within a few
hundred meters of the surface (see Figure~\ref{fig:venatm}), with
launch planned for the early 2030s.


\subsection{Instrumentation}
\label{instruments}

The DS carries five instruments, four of which are directly relevant
to exoplanet atmospheric modeling \citep{garvin2022}. The Venus Mass
Spectrometer (VMS) will conduct the first comprehensive mass
spectrometric survey of the Venus atmosphere from cloud tops to the
near-surface, measuring noble gas abundances and isotope ratios (He,
Ne, Ar, Kr, Xe), major constituents, and trace species at
parts-per-billion sensitivity. The Venus Tunable Laser Spectrometer
(VTLS) will provide high-sensitivity, species-specific measurements of
H$_2$O, HDO, SO$_2$, OCS, and CO at discrete altitude intervals
during descent, with at least one sample in the upper cloud and at
least five vertically distributed measurements below the cloud deck
\citep{garvin2022}. The Venus Atmospheric Structure Investigation
(VASI) will
measure the pressure--temperature profile and wind structure at
$\sim$10~m vertical resolution, a factor of $\geq$10 improvement over
any previous probe. The Venus Oxygen Fugacity (VfOx) experiment will
provide the first measurement of $f$O$_2$ in the deep atmosphere.
In addition to the descent sphere instruments, the Compact Ultraviolet
to Visible Imaging Spectrometer (CUVIS) on the carrier relay
spacecraft will provide UV-visible ($\sim$0.2--0.6~$\mu$m) spectral
mapping of the Venus cloud tops during flyby, characterizing the
composition and spatial distribution of the as-yet unidentified near-UV
absorber and constraining the cloud-top SO$_2$ distribution
\citep{garvin2022}.


\subsection{Projected Improvements}
\label{improvements}

\begin{deluxetable*}{lcccc}
\tablecaption{\label{tab:measurements} Atmospheric Profile
  Measurements: Pre-DAVINCI vs.\ DAVINCI.}
\tablehead{
\colhead{Parameter} &
\colhead{Pre-DAVINCI Source} &
\colhead{Pre-DAVINCI Precision} &
\colhead{Instrument} &
\colhead{DAVINCI Precision}
}
\startdata
$T(z)$ (0--70 km)          & PV/Venera  & $\pm$5--10 K; sparse            & VASI     & $\pm$0.5 K; continuous \\
$P(z)$ (0--70 km)          & PV/Venera  & $\pm$2--5\%; sparse             & VASI     & $\pm$0.1\%; continuous \\
CO$_2$ mixing ratio        & PV/Venera  & $\pm$1\%                        & VMS      & $\pm$0.1\% \\
N$_2$ mixing ratio         & PV/Venera  & $\pm$10--30\%                   & VMS      & $\pm$1\% \\
H$_2$O (0--70 km)          & Venera/RS  & factor $\sim$2--10              & VTLS/VMS & $\pm$10--20\% \\
SO$_2$ (0--70 km)          & Venera/Vega & factor $\sim$3--10              & VTLS/VMS & $\pm$10--20\% \\
CO (0--70 km)              & Remote sensing     & factor $\sim$2--3               & VTLS/VMS & $\pm$10--20\% \\
OCS (0--70 km)             & Remote sensing     & factor $\sim$2--5               & VTLS/VMS & $\pm$10--20\% \\
Noble gases                & PV MS              & $\pm$10--50\%                   & VMS      & $\pm$1--5\% \\
Noble gas isotopes         & PV MS (limited)    & $\pm$10--50\%                   & VMS      & $\pm$1--5\% \\
D/H ratio                  & PV MS              & $\pm$30\%                       & VTLS     & $\pm$2--5\% \\
$f$O$_2$ (near-surface)    & None               & Unconstrained                   & VfOx     & TBD \\
Winds (0--70 km)           & Venera/Vega & Limited                         & VASI     & Speed \& direction \\
\enddata
\tablecomments{PV = Pioneer Venus; RS = remote sensing.  Pre-DAVINCI
  sources: \citet{seiff1985}, \citet{marov2018}, \citet{sagdeev1986a},
  \citet{taylor2018}. DAVINCI projections: \citet{garvin2022}.}
\end{deluxetable*}

\begin{figure}
  \includegraphics[angle=270,width=\linewidth]{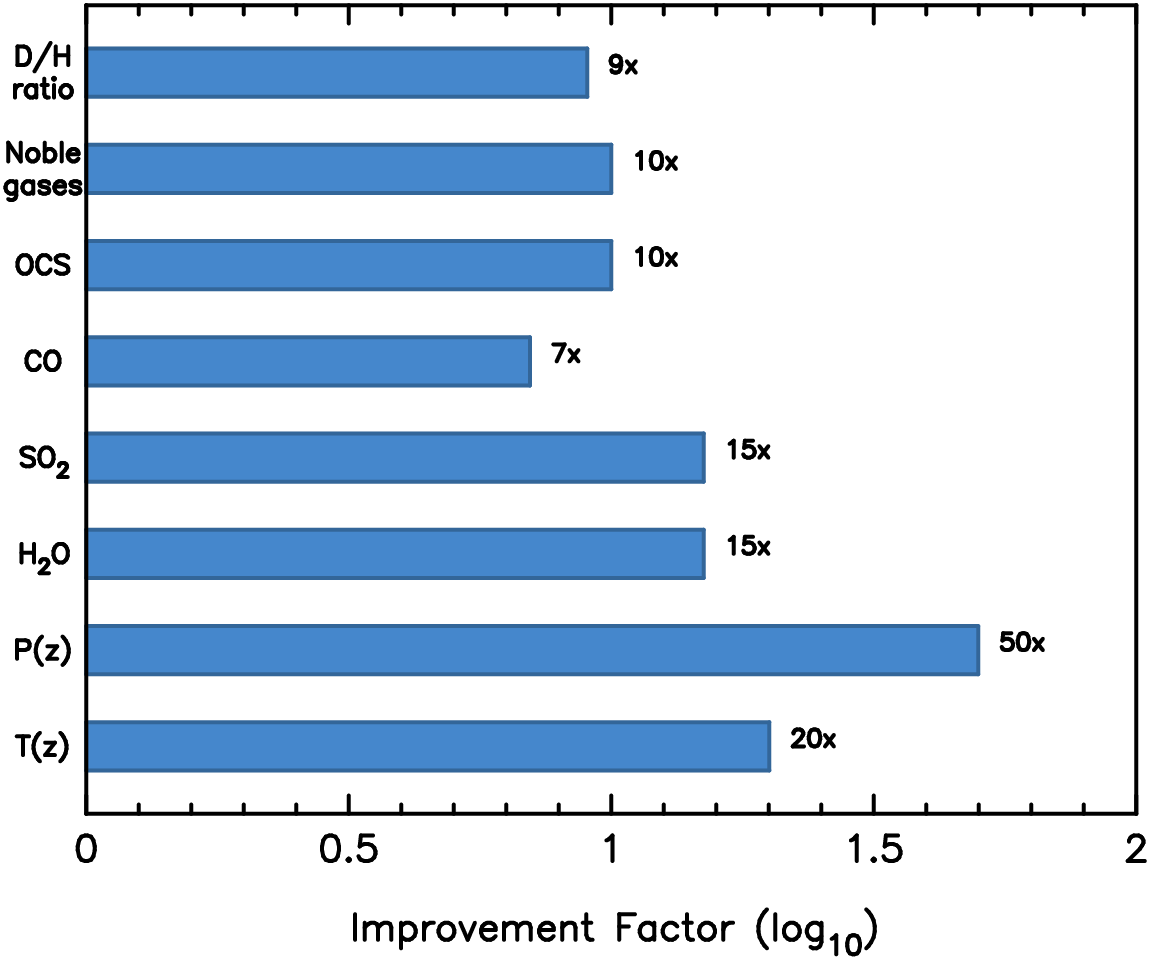}
  \caption{Improvement factors for key atmospheric profile parameters,
    defined as the ratio of pre-DAVINCI to projected DAVINCI
    measurement uncertainty, plotted on a $\log_{10}$ scale. The
    integer annotation to the right of each bar gives the improvement
    factor. Pre-DAVINCI uncertainties are from legacy in situ and
    remote sensing data
    \citep{seiff1985,marov2018,marcq2018,taylor2018}; DAVINCI
    projections are from \citet{garvin2022}. The largest gains are for
    the pressure--temperature profile ($\sim$20$\times$ and
    $\sim$50$\times$, respectively), where DAVINCI's VASI instrument
    will replace sparse, decades-old probe data with continuous
    measurements at $\sim$10~m vertical resolution. Trace gas
    improvements of $\sim$7--15$\times$ reflect the transition from
    order-of-magnitude sub-cloud ambiguities \citep{marcq2018} to
    $\pm$10--20\% precision. Noble gas and D/H ratio improvements
    ($\sim$9--10$\times$) reflect the gain from $\pm$10--50\% to
    $\pm$1--5\% precision.}
  \label{fig:venprec}
\end{figure}

Table~\ref{tab:measurements} compares key atmospheric parameters,
contrasting existing measurement precision with DAVINCI projections,
and Figure~\ref{fig:venprec} visualizes the magnitude of these
improvements. The most dramatic gains are in the pressure-temperature
profile. Pre-DAVINCI knowledge of $T(z)$ relies on Pioneer Venus and
Venera probe measurements with uncertainties of $\pm$5--10~K and
vertical sampling gaps of 1--5~km, particularly in the poorly
instrumented 12--48~km altitude range
\citep{seiff1985,marov2018}. VASI will reduce the temperature
uncertainty to $\pm$0.5~K with continuous sampling at $\sim$10~m
resolution, an improvement of $\sim$20$\times$
(Figure~\ref{fig:venprec}). The pressure profile improvement is even
larger ($\sim$50$\times$), from $\pm$2--5\% to $\pm$0.1\%. Because
both the atmospheric scale height (Equation~\ref{eq:scaleheight}) and
the amplitude of spectral features (Equation~\ref{eq:transitdepth})
depend directly on temperature, this improvement propagates into
tighter constraints on all modeled spectral features for
CO$_2$-dominated exoplanets.

The trace gas improvements, while smaller in multiplicative factor,
may be more consequential for breaking the degeneracies described in
Section~\ref{sources}. The current uncertainties in H$_2$O and SO$_2$
below the cloud deck span an order of magnitude, reflecting
discrepancies between different Venera probe results and the limited
sensitivity of remote sensing through the near-infrared spectral
windows \citep{marcq2018}. DAVINCI's VMS and VTLS will reduce these to
$\pm$10--20\% through vertically resolved measurements spanning the
full atmospheric column, providing the first reliable constraints on
how these species vary with altitude through the sub-cloud region.
These estimates are conservative; the VTLS accuracy requirements are
as low as $\sim$5\% for SO$_2$ \citep{garvin2022}, so the actual
improvement factors may be larger. The improvement factors of
$\sim$15$\times$ for
H$_2$O and SO$_2$ are critical because these species contribute
absorption features that overlap with CO$_2$ in transmission spectra,
and their unconstrained profiles are a dominant source of systematic
uncertainty in current exoplanet spectral models
\citep{lustigyaeger2019b,jordan2021}.

The noble gas and D/H ratio improvements ($\sim$10$\times$ and
$\sim$9$\times$, respectively; Figure~\ref{fig:venprec}) are
comparable in magnitude to the trace gas improvements. Although these
quantities do not directly control the transmission spectrum in the
same way as the radiatively active trace gases, the improvement in
precision for noble gas isotope ratios---from $\pm$10--50\% to
$\pm$1--5\%---will be transformative for constraining atmospheric
origin and evolution, as isotope ratios are diagnostic of volatile
delivery, hydrodynamic escape, and outgassing history
\citep{kane2021d}.


\section{Propagation to Exoplanet Spectra}
\label{sec:analysis}

The measurement improvements summarized in
Table~\ref{tab:measurements} and Figure~\ref{fig:venprec} can be
translated into improvements in the precision of the modeled Venus
benchmark spectrum through analytical error propagation. This approach
has the advantage of transparency: the contribution of each
measurement to the spectral uncertainty can be isolated and quantified
without dependence on a specific radiative transfer code configuration
or retrieval framework \citep{batalha2017b}. We derive the formalism
for transmission spectroscopy and thermal emission separately, then
evaluate the improvement factors for a fiducial Venus analog.

We emphasize that the uncertainties propagated here are those in the
modeled spectrum of Venus itself, arising from imperfect knowledge of
the Venus atmospheric profile. A measurement of the Venus temperature
at a given pressure level to $\pm$0.5~K does not imply that the
corresponding temperature on an exo-Venus is known to $\pm$0.5~K since
the exoplanet may differ in surface conditions, insolation, and
atmospheric evolution. The improvement factors we derive quantify how
much more precisely the Venus reference spectrum is known after
DAVINCI. Because Venus-like exoplanet models are built upon the Venus
atmospheric profile as a template and prior, a more precisely
characterized benchmark propagates directly into reduced systematic
uncertainty in those models. Thus, we quantify the benchmark
improvement and describe the exoplanet application as an improved
prior rather than a direct reduction in exoplanet retrieval
uncertainty.


\subsection{Transmission Spectroscopy}
\label{transmission}

The wavelength-dependent effective altitude $z_\mathrm{eff}(\lambda)$
at which the limb optical depth reaches unity determines the transit
depth. For a well-mixed absorbing species $i$ with volume mixing ratio
$x_i$ and absorption cross-section $\kappa_i(\lambda)$ in an
isothermal atmosphere with scale height $H$, the chord optical depth
through the limb at altitude $z$ is
\citep{fortney2005c,lecavelierdesetangs2008b,betremieux2015}
\begin{equation}
\tau(z,\lambda) = x_i \, \kappa_i(\lambda) \, n(z) \,
\sqrt{2 \pi R_p H}
\label{eq:chordtau}
\end{equation}
where $n(z) = n_0 \exp(-z/H)$ is the number density at altitude $z$,
$n_0$ is the number density at the reference level, and $R_p$ is the
planetary radius. Setting $\tau = 1$ and solving for
$z_\mathrm{eff}$ yields
\begin{equation}
z_\mathrm{eff}(\lambda) = H \ln \left(
\frac{x_i \, \kappa_i \, P_0 \, \sqrt{2 \pi R_p H}}
{k_B T} \right)
\label{eq:zeff}
\end{equation}
where $P_0$ is the reference pressure and $T$ is the temperature. The
transit depth is then
\begin{equation}
\delta(\lambda) = \frac{\left[ R_p +
z_\mathrm{eff}(\lambda) \right]^2}{R_\star^2}
\label{eq:delta}
\end{equation}
and the transit depth variation between an absorbing wavelength and a
nearby continuum wavelength is $\Delta\delta \approx 2 R_p \,
\Delta z_\mathrm{eff} / R_\star^2$ for $z_\mathrm{eff} \ll R_p$
\citep{seager2000b,kempton2018}.

The uncertainty in $z_\mathrm{eff}$ due to uncertainties in the
atmospheric profile parameters can be obtained by differentiating
Equation~\ref{eq:zeff}. Recalling that $H = k_B T / \mu m_u g$, the
partial derivative with respect to temperature is
\begin{equation}
\frac{\partial z_\mathrm{eff}}{\partial T} =
\frac{H}{T} \left( \ln \xi - \frac{1}{2} \right)
\label{eq:dzdT}
\end{equation}
where $\xi \equiv x_i \kappa_i P_0 \sqrt{2 \pi R_p H} / (k_B T)$
is the argument of the logarithm in Equation~\ref{eq:zeff}. Since
$z_\mathrm{eff}$ depends logarithmically on the mixing ratio $x_i$,
the natural variable for uncertainty propagation is $\ln x_i$:
\begin{equation}
\frac{\partial z_\mathrm{eff}}{\partial \ln x_i} = H
\label{eq:dzdlnx}
\end{equation}
which states that a change of one $e$-folding in the mixing ratio
shifts the effective altitude by one scale height. The total
uncertainty in the transit depth variation is then
\begin{equation}
\sigma_{\Delta\delta} = \frac{2 R_p}{R_\star^2}
\sqrt{ \left( \frac{\partial z_\mathrm{eff}}{\partial T}
\right)^2 \sigma_T^2 +
H^2 \, \sigma_{\ln x_i}^2 }
\label{eq:sigmadelta}
\end{equation}
where $\sigma_T$ is the temperature uncertainty and
$\sigma_{\ln x_i}$ is the uncertainty in the natural logarithm of the
mixing ratio. For a one-sigma multiplicative uncertainty of factor
$N$ (i.e., $x$ ranges from $x_0 / N$ to $N x_0$),
$\sigma_{\ln x} = \ln N$.

From Equation~\ref{eq:sigmadelta} it can be seen that the temperature
contribution enters through a multiplicative factor that includes $\ln
\xi$, which for a thick CO$_2$ atmosphere at the cloud-top level is of
order $\sim$10--15. The $-1/2$ term represents the decrease in
reference number density as temperature increases (since $n_0 = P_0 /
k_B T$), which partially offsets the scale height increase; for $\ln
\xi \gg 1$, this correction is small. The temperature uncertainty
therefore has an amplified effect on the spectral uncertainty. Also,
the logarithmic dependence of $z_\mathrm{eff}$ on the mixing ratio
means that even large multiplicative uncertainties in trace gas
abundances produce modest altitude shifts: a factor-of-10 uncertainty
contributes $\sigma_z = \ln(10) \, H \approx 2.3 H$, not $10 H$ as a
linear treatment would suggest. This logarithmic compression is
fundamental to transmission spectroscopy and is the reason that mixing
ratio constraints from transit observations are intrinsically coarse.


\subsection{Thermal Emission}
\label{emission}

For a directly imaged or secondary-eclipse observation, the
planet-to-star flux ratio at wavelength $\lambda$ is
\begin{equation}
\frac{F_p}{F_\star} = \left( \frac{R_p}{R_\star} \right)^2
\frac{B_\lambda(T_\mathrm{eff})}{B_\lambda(T_\star)}
\label{eq:fluxratio}
\end{equation}
where $B_\lambda(T)$ is the Planck function and $T_\mathrm{eff}$ is
the effective emission temperature at wavelength $\lambda$, which
depends on the atmospheric temperature profile and the
wavelength-dependent opacity
\citep{seager2010,kane2011g,madhusudhan2019}. The fractional
uncertainty in the flux ratio due to the temperature profile
uncertainty is
\begin{equation}
\frac{\sigma_{F_p/F_\star}}{F_p/F_\star} =
\frac{\partial \ln B_\lambda}{\partial T} \, \sigma_T
\label{eq:sigmaflux}
\end{equation}
where the Planck function sensitivity is
\begin{equation}
\frac{\partial \ln B_\lambda}{\partial T} =
\frac{h c}{\lambda \, k_B \, T^2} \,
\frac{1}{1 - \exp(-h c / \lambda k_B T)}
\label{eq:planck_sensitivity}
\end{equation}
which reduces to $1/T$ in the Rayleigh--Jeans limit ($\lambda \gg
hc/k_B T$) and to $hc / \lambda k_B T^2$ in the Wien limit
($\lambda \ll hc/k_B T$) \citep{seager2010}.

For Venus at the cloud-top emission temperature of $\sim$260~K, the
Wien limit is a good approximation at wavelengths
$\lambda \lesssim 30~\mu$m (since $hc / k_B T \approx 55~\mu$m),
and the Planck sensitivity at
$\lambda = 10~\mu$m is $\partial \ln B / \partial T \approx
0.021$~K$^{-1}$. A temperature uncertainty of $\sigma_T = 10$~K
(pre-DAVINCI) therefore produces a fractional flux uncertainty of
$\sim$21\%, while $\sigma_T = 0.5$~K (DAVINCI) reduces this to
$\sim$1\%. The thermal emission improvement is thus even larger than
the transmission improvement because of the steep dependence of the
Planck function on temperature in the Wien regime.

The deep-atmosphere temperature profile also affects thermal emission
through the greenhouse effect: the outgoing longwave radiation at each
wavelength is emitted from the altitude where the atmosphere becomes
optically thin, and the emission temperature at that altitude
determines the flux. Equation~\ref{eq:planck_sensitivity} can be
evaluated at these different emission conditions to estimate the
sensitivity across the spectrum. In the 8--12~$\mu$m atmospheric
window, thermal emission originates from deeper, warmer levels
($T \sim 350$--400~K), where the Planck sensitivity is lower
($\partial \ln B / \partial T \approx 0.009$~K$^{-1}$ at
$\lambda = 10$~$\mu$m, $T = 400$~K). With a pre-DAVINCI uncertainty
of $\sigma_T \approx 5$--10~K in this altitude range, the resulting
flux uncertainty is $\sim$5--10\%. In the 15~$\mu$m CO$_2$ band,
emission originates near the cloud top ($T \sim 260$~K), where the
Planck sensitivity is $\partial \ln B / \partial T \approx
0.015$~K$^{-1}$ and $\sigma_T = 10$~K yields $\sim$15\% flux
uncertainty. Across the mid-infrared, the pre-DAVINCI profile
uncertainties therefore propagate into flux uncertainties of
$\sim$5--15\%, depending on the emission altitude and wavelength.
DAVINCI's $\pm$0.5~K precision reduces these to $\sim$0.5--3\%. We
note that this single-layer estimate is a simplification: in a thick
atmosphere, the outgoing flux depends on the integrated opacity and
temperature structure over a range of altitudes, and a full radiative
transfer calculation would be needed to quantify the improvement
precisely. The single-layer scaling nevertheless captures the correct
order of magnitude and demonstrates that thermal emission is the
regime where DAVINCI's value is most acute.


\subsection{Improvement Factors for a Venus Analog}
\label{improvement}

We evaluate the error propagation formalism for a Venus analog
exoplanet. For the transmission case, we consider a Venus twin ($R_p =
0.95~R_\oplus$, $M_p = 0.815~M_\oplus$, $g = 8.87$~m~s$^{-2}$)
transiting an M~dwarf ($R_\star = 0.2~R_\odot$, $T_\mathrm{eff} =
3200$~K), placed at the orbital distance that reproduces the Venus
insolation ($\sim$0.044~AU for this host, corresponding to an orbital
period of $\sim$8~days), which represents the most observationally
accessible case for JWST transmission spectroscopy
\citep{kempton2018,lustigyaeger2019a,lustigyaeger2019b}. For the
thermal emission case, we also adopt an M-dwarf host, since the
planet-to-star contrast for a Venus analog around a Sun-like star is
far below the reach of current facilities. Secondary-eclipse thermal
emission is feasible only for terrestrial planets around late-type
stars with JWST/MIRI, while characterization around solar-type hosts
awaits future mid-infrared concepts such as LIFE \citep{quanz2022a}.

At the Venus cloud-top level ($T \approx 260$~K, $P \approx 0.03$~bar,
$\mu = 44$~amu), the scale height is $H \approx 5.5$~km. For the
dominant CO$_2$ absorption at 4.3~$\mu$m, the logarithmic factor
$\ln \xi \approx 12$, yielding $\partial z_\mathrm{eff} / \partial T
\approx 0.24$~km~K$^{-1}$ from Equation~\ref{eq:dzdT}. The
pre-DAVINCI temperature uncertainty of $\sigma_T = 10$~K therefore
produces an altitude uncertainty of $\sigma_z \approx 2.4$~km from
the temperature term alone, corresponding to a transit depth
uncertainty of $\sigma_{\Delta\delta} \approx 1.5$~ppm for the
M-dwarf geometry.

The trace gas contribution to the cloudy transmission spectrum depends
only on the abundances above the opacity floor, since the H$_2$SO$_4$
aerosol layers prevent transmission spectroscopy from probing below
$\sim$70~km (Section~\ref{cloudfree}). Above the cloud deck, where
Venus Express data provide constraints, the fractional mixing ratio
uncertainties are $\sigma_x / x \sim 0.3$--0.5 \citep{marcq2018}, for
which the linear and logarithmic treatments agree to within
$\sim$15\%. From Equation~\ref{eq:dzdlnx}, $\sigma_{\ln x} \approx
\sigma_x / x \approx 0.4$ produces $\sigma_z \approx 0.4 H \approx
2.2$~km, comparable to the temperature contribution. Combining both
terms in quadrature via Equation~\ref{eq:sigmadelta} yields a total
cloud-top transit depth uncertainty of $\sigma_{\Delta\delta} \approx
2$--4~ppm pre-DAVINCI.  With DAVINCI ($\sigma_T = 0.5$~K, $\sigma_x /
x \sim 0.15$), the temperature contribution drops to $\sigma_z \approx
0.12$~km and the trace gas contribution to $\sigma_z \approx 0.8$~km,
yielding a combined $\sigma_{\Delta\delta} \approx 0.5$~ppm, which is
an improvement of $\sim$4--5$\times$.

The sub-cloud trace gas abundances do not directly affect the cloudy
transmission spectrum, but they become relevant in two cases.  First,
in the cloud-free limit (Section~\ref{cloudfree}), the opacity floor
drops to $\sim$35~km and the sub-cloud region becomes
spectroscopically accessible. Second, thermal emission and the deep
atmospheric structure depend on the sub-cloud composition through the
temperature profile and the greenhouse effect. For these cases, the
sub-cloud uncertainties are substantial and the logarithmic
propagation becomes essential. Consider SO$_2$ with a sub-cloud mixing
ratio of $x \sim 10^{-3.8}$ and a pre-DAVINCI uncertainty spanning an
order of magnitude \citep{marcq2018}. An order-of-magnitude
uncertainty corresponds to $\sigma_{\ln x} = \ln(10) \approx 2.3$,
yielding $\sigma_z \approx 2.3 H \approx 13$~km from
Equation~\ref{eq:dzdlnx}. In the cloud-free case, this is comparable
to the observable atmospheric column above the $\sim$35~km floor,
meaning that the sub-cloud SO$_2$ contribution is effectively
unconstrained by current data. With DAVINCI ($\sigma_x / x \sim 0.15$,
i.e., $\sigma_{\ln x} \approx 0.15$), this reduces to $\sigma_z
\approx 0.8$~km, an improvement of $\sim$15$\times$.

\begin{figure*}
  \includegraphics[angle=270,width=\textwidth]{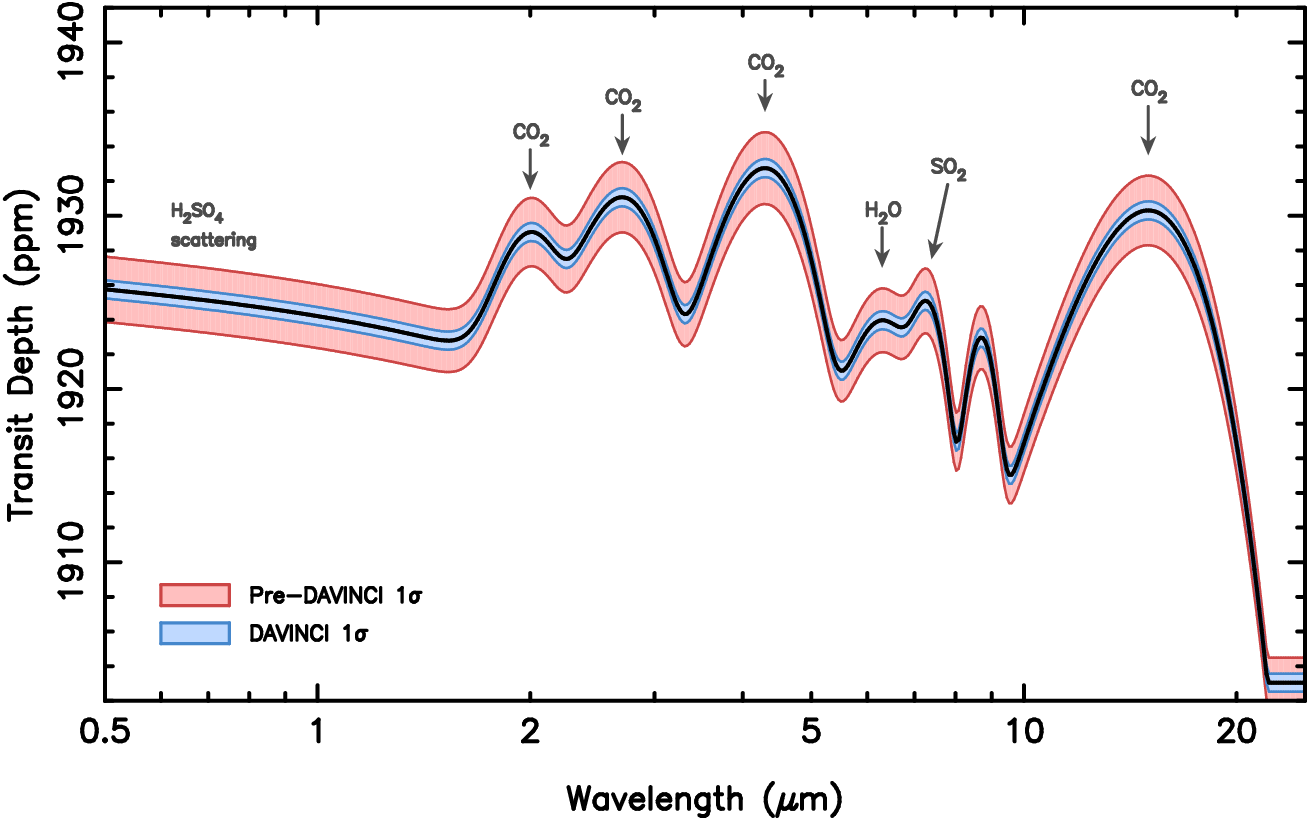}
  \caption{Illustrative transit depth and analytical uncertainty
    envelopes for a Venus twin ($R_p = 0.95~R_\oplus$) transiting an
    M4V host ($R_\star = 0.2~R_\odot$). The black curve shows the
    baseline transit depth computed from an illustrative opacity model
    with Gaussian absorption band profiles for CO$_2$ (2.0, 2.7, 4.3,
    and 15~$\mu$m), H$_2$O (6.3~$\mu$m), and SO$_2$ (7.3 and
    8.7~$\mu$m), plus H$_2$SO$_4$ haze and CO$_2$ Rayleigh
    scattering. The red envelope shows the $1\sigma$ model systematic
    from pre-DAVINCI atmospheric profile knowledge, computed from
    Equations~\ref{eq:dzdT}--\ref{eq:sigmadelta} with $\sigma_T =
    10$~K and above-cloud trace gas uncertainties of $\sigma_{\ln x}
    \approx 0.4$, appropriate to the cloudy transmission case in which
    only abundances above the $\sim$70~km opacity floor are
    probed. The blue envelope shows the DAVINCI-calibrated uncertainty
    ($\sigma_T = 0.5$~K, $\sigma_{\ln x} \approx 0.15$). The wider red
    pre-DAVINCI envelope ($\pm$2~ppm) and the narrower blue DAVINCI
    envelope ($\pm$0.5~ppm) are both visible around the black
    baseline, illustrating the factor of $\sim$4 reduction in the
    modeled benchmark uncertainty for the cloudy transmission case.}
    \label{fig:ventran}
\end{figure*}

Figure~\ref{fig:ventran} shows the resulting transit depth and
uncertainty envelopes as a function of wavelength, computed from the
analytical formalism for the fiducial M-dwarf transit
case. Table~\ref{tab:improvement} summarizes the combined improvement
factors for the key observable quantities. As shown in
Figure~\ref{fig:ventran}, the overall improvement in transit depth
precision is a factor of $\sim$4--5, driven by the combined
contributions of the temperature and trace gas improvements. The
thermal emission improvement is larger ($\sim$5--15$\times$) because
of the steep Planck function sensitivity to temperature.

\begin{figure}
  \includegraphics[angle=270,width=\columnwidth]{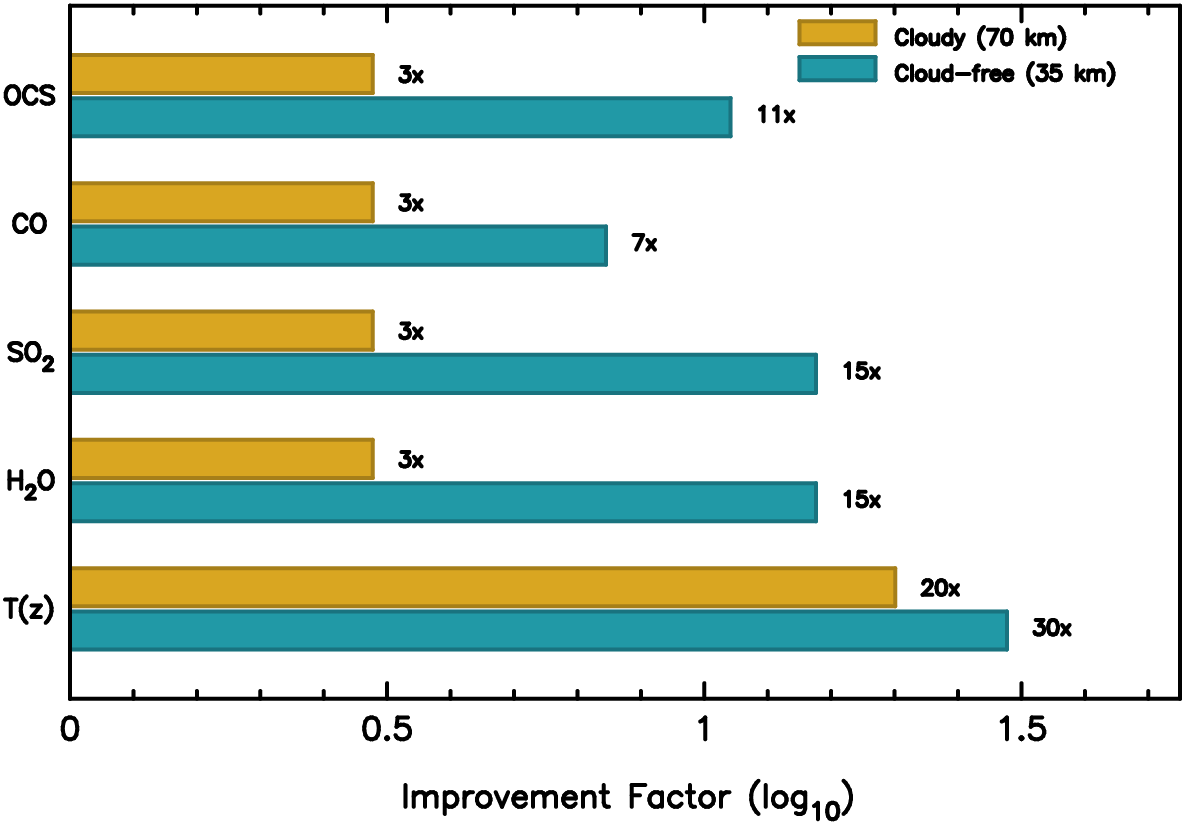}
  \caption{DAVINCI improvement factors for key atmospheric parameters,
    evaluated at the cloudy opacity floor (70~km, $\sim$0.03~bar; gold
    bars) and the cloud-free opacity floor (35~km, $\sim$5~bar; teal
    bars), using the uncertainties appropriate to each altitude level.
    At the cloudy floor, the above-cloud trace gas uncertainties are
    modest ($\sigma_{\ln x} \sim 0.4$), so the temperature improvement
    dominates. At the cloud-free floor, the sub-cloud trace gas
    uncertainties are much larger ($\sigma_{\ln x} \sim 1.6$--2.3,
    corresponding to factor-of-5 to order-of-magnitude ambiguities),
    and the DAVINCI improvement in trace gas constraints becomes the
    dominant contribution. This demonstrates that DAVINCI's value for
    constraining exoplanet atmospheric models is robust regardless of
    cloud properties, and is most consequential for the sub-cloud
    atmospheric region where current data are poorest.
    \label{fig:venimp}}
\end{figure}

The above analysis assumes the cloudy opacity floor at $\sim$70~km
($\sim$0.03~bar). As discussed in Section~\ref{cloudfree}, a
cloud-free CO$_2$ atmosphere would lower the floor to $\sim$35~km
($\sim$5~bar) due to Rayleigh scattering, CIA, and refraction. The
error propagation formalism can be evaluated at this deeper level to
assess how the improvement factors change. At 35~km, the VIRA
temperature is $\sim$490~K \citep{seiff1985}, yielding a scale height
of $H \approx 10$~km, roughly twice the cloud-top value. The
pre-DAVINCI uncertainties are worse in this altitude range because the
sub-cloud region (12--48~km) is where the legacy in situ data are
sparsest and most discrepant: temperature uncertainties may reach
$\pm$15~K, and H$_2$O and SO$_2$ mixing ratios are uncertain by a full
order of magnitude \citep{marcq2018}. At this deeper pressure level,
the increased reference density ($n_0 \propto P/T$) raises the
logarithmic optical-depth factor to $\ln \xi \sim 20$ (compared to
$\sim$12 at the cloud tops), and the larger scale height yields
$\partial z_\mathrm{eff} / \partial T \approx 0.40$~km~K$^{-1}$ from
Equation~\ref{eq:dzdT}. The pre-DAVINCI altitude uncertainty therefore
rises to $\sigma_z \approx 6$~km.  DAVINCI reduces this to $\sigma_z
\approx 0.20$~km, an improvement of $\sim$30$\times$ for the
temperature term. For trace gas species, the order-of-magnitude
sub-cloud uncertainties yield $\sigma_z = \ln(10) \, H \approx 2.3
\times 10 \approx 23$~km from Equation~\ref{eq:dzdlnx}, while DAVINCI
reduces this to $\sigma_z \approx 1.5$~km, an improvement of
$\sim$15$\times$. The DAVINCI improvement factors are therefore larger
at the cloud-free opacity floor than at the cloud tops, precisely
because the pre-DAVINCI data are worst in the sub-cloud region that
becomes newly accessible without aerosols. This strengthens the case
for DAVINCI's value regardless of the cloud properties of a given
exoplanet: whether the opacity floor is set by aerosols at 70~km or by
gas-phase processes at 35~km, DAVINCI provides the critical ground
truth for the altitude range that dominates the spectral
uncertainty. Figure~\ref{fig:venimp} summarizes the improvement
factors at both opacity floors for the key atmospheric parameters,
illustrating that the DAVINCI value proposition is consistently
stronger at the cloud-free floor.

\begin{deluxetable}{lccc}
\tablecaption{\label{tab:improvement} Spectral Uncertainty
  Improvement Factors from DAVINCI.}
\tablehead{
\colhead{Observable} &
\colhead{Pre-DAVINCI} &
\colhead{DAVINCI} &
\colhead{Factor}
}
\startdata
$\sigma_{\Delta\delta}$ at 4.3~$\mu$m (ppm) & $\sim$2 & $\sim$0.5 & 4$\times$ \\
$\sigma_{\Delta\delta}$ at 2.7~$\mu$m (ppm) & $\sim$2 & $\sim$0.5 & 4$\times$ \\
$\sigma_{\Delta\delta}$ at 15~$\mu$m (ppm)  & $\sim$3 & $\sim$0.5 & 6$\times$ \\
$\sigma_{F_p/F_\star}$ at 10~$\mu$m (\%)    & $\sim$15 & $\sim$1 & 15$\times$ \\
$\sigma_{F_p/F_\star}$ at 15~$\mu$m (\%)    & $\sim$10 & $\sim$2 & 5$\times$ \\
$\sigma_{\log x}$(H$_2$O)                   & $\sim$1.0 dex & $\sim$0.07 dex & 15$\times$ \\
$\sigma_{\log x}$(SO$_2$)                   & $\sim$1.0 dex & $\sim$0.07 dex & 15$\times$ \\
$\sigma_{\log x}$(CO)                       & $\sim$0.5 dex & $\sim$0.07 dex & 7$\times$ \\
$\sigma_{\log x}$(OCS)                      & $\sim$0.7 dex & $\sim$0.07 dex & 10$\times$ \\
\enddata
\tablecomments{Transit depth uncertainties
  ($\sigma_{\Delta\delta}$) are cloud-top values evaluated for a Venus
  twin transiting an M4V host using Equation~\ref{eq:sigmadelta} with
  $\sigma_T = 10$~K and $\sigma_{\ln x} \approx 0.4$ (see
  Section~\ref{improvement}). These are cloudy-case values, for
  which only above-cloud abundances are probed. In the cloud-free
  limit the sub-cloud region becomes accessible and the improvement
  factors are larger (Section~\ref{improvement},
  Figure~\ref{fig:venimp}). Thermal emission uncertainties
  ($\sigma_{F_p/F_\star}$) are evaluated for a Venus twin
  transiting an M4V host, since thermal emission characterization of a
  Venus analog around a Sun-like star is beyond the reach of current
  facilities. Trace gas uncertainties
  ($\sigma_{\log x}$) reflect the sub-cloud measurement precision from
  Table~\ref{tab:measurements}, expressed in dex; pre-DAVINCI values
  correspond to the factor-of-3 to factor-of-10 ambiguities reported
  by \citet{marcq2018}, and DAVINCI values correspond to
  $\pm$10--20\% precision. Surface temperature and pressure cannot be
  directly constrained from transit or emission observations and are
  not listed; constraining these quantities for exoplanets is one of
  the primary motivations for Venus ground truth (see
  Section~\ref{sources}).}
\end{deluxetable}


\subsection{Resolving the Degeneracies}
\label{resolving}

The analytical framework and improvement factors presented above can
now be connected explicitly to the four sources of degeneracy
identified in Section~\ref{sources}.

The {\em cloud-composition degeneracy} arises because cloud properties
and gas-phase abundances produce partially interchangeable spectral
effects. DAVINCI will break this degeneracy by measuring both
independently and simultaneously during descent through the cloud
layer. The VMS and VTLS will measure the gas-phase composition above,
within, and below the cloud deck, while VASI's pressure, temperature,
and accelerometer data will constrain the cloud vertical extent, base
and top altitudes, and density structure. Together, these provide the
cloud optical depth profile and the gas-phase absorption
independently, eliminating the ambiguity that arises when only the
combined spectral effect is observed remotely.

The {\em pressure-temperature degeneracy} is addressed most
directly. DAVINCI's VASI instrument will replace the sparse,
decades-old $T$--$P$ profiles from Pioneer Venus and Venera with
continuous measurements at $\sim$10~m vertical resolution and
$\pm$0.5~K precision from the cloud tops to the near-surface
\citep{garvin2022}. As shown in Section~\ref{improvement}, this
reduces the temperature-driven transit depth uncertainty by
$\sim$20$\times$ at the cloud-top level and $\sim$30$\times$ at the
cloud-free opacity floor, and reduces thermal emission flux
uncertainties from $\sim$5--15\% to $\sim$0.5--3\%
(Table~\ref{tab:improvement}). By providing the full lapse rate
structure through the sub-cloud region, DAVINCI will break the
ambiguity between high surface pressure with a steep lapse rate and
lower surface pressure with a shallower lapse rate that currently
permits a wide range of surface conditions to be consistent with the
same cloud-top observations.

The {\em trace gas profile degeneracy} is resolved through the
combination of VMS and VTLS, which will provide vertically resolved
measurements of H$_2$O, SO$_2$, CO, and OCS from the upper cloud
deck to the near-surface, with VMS additionally measuring HCl and HF
\citep{garvin2022}. The current order-of-magnitude ambiguities in
sub-cloud trace gas abundances \citep{marcq2018} produce altitude
uncertainties of $\sim$2$H$ in the effective transmission altitude
(Section~\ref{improvement}). DAVINCI reduces these to
$\sim$0.15$H$, an improvement of $\sim$15$\times$
(Figure~\ref{fig:venimp}). These measurements will reveal the
vertical gradients driven by photochemistry, thermochemistry, and
heterogeneous cloud reactions that are currently interpolated or
assumed in exoplanet atmospheric models.

The {\em surface-atmosphere interaction degeneracy} is the most
challenging because it involves the planetary surface, which DAVINCI
does not directly image. However, DAVINCI addresses it from the
atmospheric side through two measurements. First, the VMS and VTLS
near-surface trace gas measurements will reveal whether species such
as SO$_2$, OCS, and CO are in thermochemical equilibrium with the
surface, constraining the surface mineralogy and redox state
indirectly \citep{fegley1997a,zolotov2018b}. Second, the VfOx
instrument will provide the first direct measurement of oxygen
fugacity ($f$O$_2$) in the deep Venus atmosphere, a quantity that is
diagnostic of the redox buffering capacity of the near-surface
environment \citep{garvin2022}. When combined with VERITAS surface
emissivity measurements (Section~\ref{synergy}), these data will
provide the most complete characterization of surface--atmosphere
equilibrium on a CO$_2$-dominated terrestrial planet, establishing the
ground truth needed to constrain the bottom boundary condition in
exoplanet atmospheric models.


\section{Discussion}
\label{sec:disc}


\subsection{From DAVINCI Data to Exoplanet Models}
\label{application}

The improvement factors quantified in Section~\ref{sec:analysis} will
only be realized if the DAVINCI atmospheric profile data are
accessible to the exoplanet modeling community in suitable formats.
The DAVINCI descent sphere will deliver a self-consistent dataset
comprising the complete $T$-$P$-$\rho$ profile at $\sim$10~m vertical
resolution from VASI, vertically resolved trace gas measurements from
VMS and VTLS at discrete altitude intervals through the full
atmospheric column, and the near-surface $f$O$_2$ measurement from
VfOx. These data products will serve two distinct functions for
exoplanet science. First, they will provide a validated a priori
atmospheric profile that can replace the ad hoc profiles currently
assembled from heterogeneous historical data as inputs to forward
models such as PSG \citep{villanueva2018b}, SMART \citep{meadows1996},
and petitRADTRANS \citep{molliere2019b}. Second, they will serve as a
validation benchmark for atmospheric retrieval algorithms: any
retrieval code applied to the DAVINCI Venus data should recover the
known atmospheric properties, and failures to do so would identify
systematic biases in the retrieval methodology. This validation
approach is analogous to the use of Earth's transmission spectrum as a
retrieval benchmark \citep{lustigyaeger2023a}, but Venus provides a
more demanding test given its thick cloud layer and deep, hot
atmosphere, precisely the conditions that will be encountered in
characterizing CO$_2$-dominated exoplanet atmospheres.

Beyond direct application to Venus analogs, the DAVINCI data will
enable parameterized atmospheric models that scale to different
surface pressures, temperatures, stellar irradiation levels, and
planetary masses, supporting robust predictions for VZ planets around
different stellar types \citep{kane2014e,ostberg2023a,miles2025}. The
noble gas measurements will constrain atmospheric origin and evolution
relevant to understanding the diversity of CO$_2$-dominated
atmospheres \citep{kane2021d}. The combination of high-resolution
profile data with the analytical framework developed in
Section~\ref{sec:analysis} will allow the exoplanet community to
quantify, for the first time, how well a given set of transit or
emission observations can constrain the atmospheric structure of a
Venus-like world.


\subsection{Venus Zone Demographics}
\label{demographics}

The VZ framework \citep{kane2014e} identifies the orbital region where
a terrestrial planet is likely to experience a runaway greenhouse.
Demographic analyses of the Kepler and TESS catalogs indicate that VZ
terrestrial planets are at least as common as HZ terrestrial planets
\citep{kane2016c,hill2023,ostberg2023a}, and their characterization is
a central objective for JWST and HWO \citep{kane2024b,kane2026a}. The
degeneracies described here are a significant obstacle since, without
ground truth for a thick CO$_2$-dominated atmosphere, VZ exoplanet
spectra will remain systematics-limited. Even the detection of CO$_2$
in a terrestrial exoplanet atmosphere is insufficient to distinguish a
Venus-like from a thin-atmosphere scenario
\citep{lustigyaeger2019b,ostberg2023c}. DAVINCI provides the
calibration needed to advance beyond this limitation.


\subsection{Synergy with Other Venus Missions}
\label{synergy}

The exoplanet value of DAVINCI is amplified by its concurrent
operation with two other Venus missions, each of which addresses
complementary aspects of the degeneracies described in
Section~\ref{sec:degen}.

VERITAS will provide global surface emissivity measurements at near-
infrared wavelengths using the Venus Emissivity Mapper (VEM), enabling
the first global constraints on surface mineralogy and rock type
\citep{smrekar2023}. For the exoplanet application, this is critical
because the surface--atmosphere interaction degeneracy
(Section~\ref{sources}) depends on the mineralogy and redox state of
the surface, which controls weathering and outgassing reactions that
buffer the atmospheric composition
\citep{fegley1997a,gillmann2022}. DAVINCI's near-surface
chemistry measurements probe this interaction from the atmospheric
side, while VERITAS surface emissivity constrains it from the surface
side. Together, they will provide the first self-consistent
characterization of a surface--atmosphere equilibrium on a
CO$_2$-dominated terrestrial planet, which is the ground truth needed
to interpret the bottom boundary condition in exoplanet atmospheric
models.

ESA's EnVision mission will contribute high-resolution radar mapping,
subsurface sounding, and spectroscopic monitoring of the atmosphere
and cloud layer \citep{widemann2023}. The atmospheric observations
are particularly relevant: EnVision will characterize the spatial and
temporal variability of SO$_2$, H$_2$O, and cloud properties above
the cloud deck, providing the context needed to assess how
representative DAVINCI's single-location profile is of the global
mean atmosphere. This addresses a key limitation noted in
Section~\ref{limitations}: DAVINCI provides one profile at one
location and local time, and the degree to which it represents the
globally-averaged atmosphere must be assessed using concurrent remote
sensing. \citet{way2023a} provide a comprehensive discussion of the
synergies between Venus missions and exoplanetary science, and
emphasize that the combination of in situ, orbital, and remote sensing
datasets is essential for constructing the multi-parameter Venus
characterization that exoplanet models require.


\subsection{Implications for Future Direct Imaging Missions}
\label{hwo}

HWO will characterize terrestrial exoplanets via reflected-light
spectroscopy at UV/optical/near-infrared wavelengths ($\sim$0.2--2~
$\mu$m), with an expected yield of $\sim$25 potentially habitable
worlds \citep{stark2024b,tuchow2025a}. A significant fraction of the
HWO target list may host planets in or near the VZ
\citep{kane2024d,harada2025}, and the ability to confidently
distinguish Venus analogs from potentially habitable worlds is
essential for efficient use of observing time
\citep{kane2026a}. In reflected light, Venus analogs present
distinctive spectral signatures: high broadband albedo from the thick
H$_2$SO$_4$ cloud deck, strong UV absorption from the as-yet
unidentified near-UV absorber, and molecular features from CO$_2$ and
H$_2$O in the near-infrared that depend on the cloud-top altitude and
above-cloud composition
\citep{arney2014,damiano2022a,ostberg2023c}. Accurate modeling of
these reflected-light signatures requires knowledge of the cloud
vertical structure, microphysical properties, and above-cloud gas
composition, all of which DAVINCI will measure in situ for the first
time with modern instrumentation. The CUVIS instrument will
additionally provide the first detailed UV-visible spectral
characterization of the near-UV absorber from the carrier spacecraft,
directly constraining the spectral feature that is one of the most
distinctive reflected-light diagnostics of a Venus-like atmosphere.

The DAVINCI atmospheric profile will also serve as an anchor point for
spectral retrievals of reflected-light
observations. \citet{damiano2022a} demonstrated that reflected-light
retrievals for terrestrial exoplanets can constrain atmospheric
composition and cloud properties, but the accuracy of these retrievals
depends on the fidelity of the forward model
atmosphere. DAVINCI-calibrated atmospheric profiles will enable the
construction of reflected-light spectral templates against which HWO
observations can be compared. For thermal emission characterization,
future mid-infrared concepts such as the Large Interferometer for
Exoplanets (LIFE; \citealt{quanz2022a,quanz2022b}) would directly
probe the 3--20~$\mu$m wavelength range where the DAVINCI-calibrated
improvement factors are largest (Table~\ref{tab:improvement}). In this
regime, the $\sim$5--15\% pre-DAVINCI model systematic would
fundamentally limit the scientific return of any mid-infrared
characterization mission, making DAVINCI ground truth essential for
the design and interpretation of such observations.


\subsection{Limitations}
\label{limitations}

An important caveat to the improvement factors derived in
Section~\ref{sec:analysis} is that they quantify the reduction in
model systematic uncertainty, not the total observational
uncertainty. The degree to which DAVINCI-calibrated models improve the
interpretation of actual exoplanet observations depends on where the
systematics-vs-photon-noise boundary falls for a given target and
instrument configuration.

For JWST transmission spectroscopy of a Venus analog transiting an M4V
host, the per-transit photon noise at $R = 100$ is typically
$\sim$50--100~ppm per spectral element
\citep{batalha2018b,lustigyaeger2019a}. With $\sim$10--20 co-added
transits, achievable for favorable targets over the JWST lifetime, the
noise floor drops to $\sim$10--20~ppm, which remains larger than the
pre-DAVINCI cloud-top model systematic of $\sim$2--3~ppm
(Table~\ref{tab:improvement}). In this regime, the observational noise
dominates and the DAVINCI improvement in transit depth precision is
not the limiting factor. For the cloud-free case, where the opacity
floor drops to $\sim$35~km, the sub-cloud region becomes accessible
and the DAVINCI improvement in transmission is correspondingly larger.

The situation is qualitatively different for thermal emission. For
secondary-eclipse observations with JWST/MIRI or future mid-infrared
concepts, the $\sim$5--15\% model systematic in the flux ratio
(Table~\ref{tab:improvement}) would dominate over the photon noise for
favorable targets, making the observations fundamentally
systematics-limited. The DAVINCI-calibrated improvement to
$\sim$0.5--3\% would be essential for constraining the thermal
structure of Venus-like atmospheres in this regime.

The analytical framework presented here also has intrinsic
limitations. It assumes an isothermal, well-mixed atmosphere in the
derivation of the chord optical depth (Equation~\ref{eq:chordtau}),
whereas the real Venus atmosphere has steep temperature gradients
\citep{seiff1985,lebonnois2010} and altitude-dependent composition
\citep{krasnopolsky2012a,bierson2020}. The error propagation treats
uncertainties as independent and Gaussian, which is a simplification:
the temperature and composition profiles are physically coupled
through photochemical and thermochemical equilibria
\citep{krasnopolsky2012a}, and the pre-DAVINCI uncertainties are
better characterized as bounded ranges than Gaussian distributions
\citep{marcq2018}. A full numerical treatment using radiative transfer
codes such as PSG \citep{villanueva2018b} or petitRADTRANS
\citep{molliere2019b} with Monte Carlo sampling of the profile
uncertainties \citep{batalha2017b} would provide more rigorous
spectral uncertainty estimates, and represents a natural follow-up
investigation once DAVINCI data are in hand.

DAVINCI will provide a single profile at one location (Alpha Regio)
and local time (approximately noon). Spatial and temporal variability,
particularly in the cloud layer
\citep{jordan2021,widemann2023}, will need to be assessed using CRIS
remote sensing and complementary observations from EnVision
(Section~\ref{synergy}). The translation from Venus to exoplanet
models requires scaling assumptions; DAVINCI provides one anchor
point, and additional solar system bodies---Earth
\citep{lustigyaeger2023a,robinson2011a}, Mars, and Titan---together
with theoretical modeling will be needed for robust scaling
relationships \citep{kane2019d,kane2021d,way2023a}.


\section{Conclusions}
\label{sec:con}

The characterization of terrestrial exoplanet atmospheres with JWST
and HWO is limited by degeneracies arising from the inaccessibility of
the deep atmosphere and surface conditions in thick, cloud-covered
CO$_2$-dominated worlds. Venus, as the only accessible analog,
provides the indispensable ground truth, but current knowledge relies
on decades-old in situ data with known limitations.

DAVINCI will transform this situation by providing the first
comprehensive, high-precision atmospheric profile from cloud tops to
the near-surface, with continuous $T$--$P$ structure from VASI and
vertically resolved composition measurements from VMS and
VTLS. Through an analytical error propagation framework applied to the
transmission and thermal emission formalisms, we have demonstrated
that the spread in the modeled Venus benchmark spectrum arising from
current profile uncertainties will decrease by factors of $\sim$4--5
for transmission and $\sim$5--15$\times$ for thermal emission after
DAVINCI (Table~\ref{tab:improvement}), providing a correspondingly
sharper benchmark and prior for the modeling of exoplanets with
Venus-like atmospheres. The improvement is driven by the combined
effect of the order-of-magnitude gain in temperature precision, which
propagates through both the atmospheric scale height and the Planck
function, and the reduction of trace gas uncertainties from
order-of-magnitude ambiguities to $\pm$10--20\%, corresponding to
improvement factors of $\sim$7--15$\times$ in sub-cloud mixing ratio
constraints. Importantly, the improvement factors are larger at the
cloud-free opacity floor ($\sim$35~km) than at the cloudy floor
($\sim$70~km; Figure~\ref{fig:venimp}), because the pre-DAVINCI data
are most discrepant in the sub-cloud region that becomes newly
accessible without aerosols. DAVINCI's value for exoplanet atmospheric
modeling is therefore robust regardless of whether a given
CO$_2$-dominated world possesses Venus-like cloud layers. These
improvements address all four sources of spectral degeneracy
identified in this work: the cloud-composition, pressure-temperature,
trace gas profile, and surface-atmosphere interaction degeneracies
that currently prevent the reliable interpretation of thick,
CO$_2$-dominated exoplanet atmospheres.

The DAVINCI atmospheric profile will provide the exoplanet community
with a validated, high-resolution template for CO$_2$-dominated
atmospheres that can serve as both an a priori input for forward
models and a benchmark for retrieval algorithm validation. This will
enable more reliable interpretation of JWST observations of VZ
exoplanets and inform target selection and observing strategies for
HWO. Venus is not merely a cautionary tale in the search for habitable
worlds; it is the anchor point for decoding the atmospheres of
CO$_2$-dominated terrestrial planets throughout the galaxy
\citep{kane2024b}. DAVINCI will provide the empirical foundation for
that effort, representing a critical bridge between solar system
science and the characterization of terrestrial exoplanet atmospheres
in the coming decades.


\section*{Acknowledgements}

The authors would like to thank the anonymous referee, whose feedback
helped to improve the manuscript. The authors acknowledge funding
support from the NASA Discovery Program for the DAVINCI Science
Team. This research has made use of the NASA Exoplanet Archive, which
is operated by the California Institute of Technology, under contract
with the National Aeronautics and Space Administration under the
Exoplanet Exploration Program. The results reported herein benefited
from collaborations and/or information exchange within NASA's Nexus
for Exoplanet System Science (NExSS) research coordination network
sponsored by NASA's Science Mission Directorate. The Jet Propulsion
Laboratory is operated by the California Institute of Technology under
contract with the National Aeronautics and Space Administration
(80NM0018D0004).




\end{document}